\documentclass[aps,prb,twocolumn,showpacs,superscriptaddress]{revtex4}

\usepackage{graphicx}
\usepackage{bm}
\usepackage{amsmath}
\usepackage{dcolumn}%

\newcommand{\dxy}{\ensuremath{d_{xy}}} 
\newcommand{\dxz}{\ensuremath{d_{xz}}} 
\newcommand{\dyz}{\ensuremath{d_{yz}}} 
\newcommand{\dzz}{\ensuremath{d_{3z^2-1}}} 
\newcommand{\dxxyy}{\ensuremath{d_{x^2-y^2}}}

\newcommand{\ef}{\ensuremath{\varepsilon_{F}}}

\newcolumntype{/}{D{/}{/}{2,2}}  
\newcolumntype{.}{D{.}{.}{0}}  
 
\begin{document} 
 
\title{Electronic structure, phonon spectra and electron-phonon interaction in
  HfB$_2$}
 
\author{S.M. Sichkar} 
\affiliation{Institute of Metal Physics, 36 Vernadsky Street, 03142 
Kiev, Ukraine} 
 
\author{V.N. Antonov} 
 
\affiliation{Institute of Metal Physics, 36 Vernadsky Street, 03142 
Kiev, Ukraine}

\date{\today} 
 
\begin{abstract} 
 
The electronic structure, Fermi surface, angle dependence of the cyclotron
masses and extremal cross sections of the Fermi surface, phonon spectra,
electron-phonon Eliashberg and transport spectral functions, temperature
dependence of electrical resistivity of the HfB$_2$ diboride were investigated
from first principles using the fully relativistic and full potential linear
muffin-tin orbital methods. The calculations of the dynamic matrix were
carried out within the framework of the linear response theory. A good
agreement with experimental data of electron-phonon spectral functions,
electrical resistivity, cyclotron masses and extremal cross sections of the
Fermi surface was achieved.
 
\end{abstract} 
 
 
\maketitle

\section{\label{sec:introd}Introduction} 
 
Ceramics based on transition metal borides, nitrides, and carbides have
extremely high melting points ($>$2500 $^{\circ}$C) and are referred to as
ultra-high temperature ceramics. \cite{UYH97,FHT+07} Among them, diborides
such as ZrB$_2$ and HfB$_2$ have a unique combination of mechanical and
physical properties: high melting points ($>$3000 $^{\circ}$C); high thermal
and electrical conductivity; chemical inertness against molten metals; great
thermal shock resistance.  \cite{UYH97,FHT+07,Mroz94} Thus, although carbides
typically have the highest melting points ($>$3500 $^{\circ}$C), the diborides
ZrB$_2$ and HfB$_2$ are more attractive candidates for high-temperature
thermomechanical structural applications at temperatures $\geq$3000
$^{\circ}$C. \cite{UYH97,FHT+07} Potential applications include thermal
protective structures for leading-edge parts on hypersonic re-entry space
vehicles, \cite{UYH97,Brown97} propulsion systems, \cite{UYH97,Brown97}
furnace elements, \cite{Kuw02} refractory crucibles, \cite{Kuw02} and
plasma-arc electrodes. \cite{Kuw02,NEB+99} 
 
The discovery of superconductivity in MgB$_2$ at 39 K by Akimitsu
\cite{NNM+01} has lead to booming activity in the physics community and
activated a search for superconductivity in other diborides.  Natural
candidates for this search are AB$_2$-type light metal diborides (A = Li, Be,
Al). However, up to now superconductivity has not been reported in the
majority of these compounds. \cite{GSZ+01} Only very recently has
superconductivity below 1 K ($T_c$ = 0.72 K) been reported in
BeB$_{2.75}$. \cite{cm:YAC+01} According to Ref. \onlinecite{BuYa01} no
superconducting transition down to 0.42 K has been observed in powders of
diborides of transition metals (A = Ti, Zr, Hf, V, Ta, Cr, Mo, U). Only
NbB$_2$ is expected to superconduct with a rather low transition temperature
($<$ 1 K), and contradictory reports about superconductivity up to $T_c$=9.5 K
in TaB$_2$ can be found in Ref. \onlinecite{BuYa01}. Finally, the reported
$T_c$=7 K in ZrB$_2$ encourages further studies of these diborides.
\cite{GSZ+01}
 
Presently, a number of experimental studies exist dealing with the physical
properties of HfB$_2$ such as thermal and electrical properties,
\cite{WOCB+04,ZRDZ+06,ZPMG11,MKRM12} mechanical, \cite{DGPF+11} and elastic
properties, \cite{WMH69} the de Haas-van Alphen (dHvA) measurements of the
Fermi surface, \cite{PSD+07} optical ellipsometry measurements, \cite{YJSK+07}
magnetic susceptibility, \cite{GFL+09,FGP+09} and NMR measurements.
\cite{LuLa05} First-principles calculations of the electronic structure of
diborides including HfB$_2$ have been also presented.
\cite{VRR+01,ZLH+08,GFL+09,FGP+09,ZC10,ZLLH+10,DCC10a,ZCL11,LZZC11,LBD11}

Lawson {\it et al.} \cite{LBD11} studied the electronic structure and
lattice properties of HfB$_2$ and ZrB$_2$ in a frame of the density
functional theory (DFT). Lattice constants and elastic constants were
determined. Computations of the electronic density of states, band
structure, electron localization function, etc. show the diverse
bonding types that exist in these materials. They also suggest the
connection between the electronic structure and the superior
mechanical properties. Lattice dynamical effects were considered,
including phonon dispersions, vibrational densities of states, and
specific heat curves. Point defect (vacancies and antisites)
structures and energetics are also presented. Vajeeston {\it et al.}
\cite{VRR+01} investigated the electronic structure of HfB$_2$ using
the tight-bonding linear muffin-tin orbital method, they claimed that
metal-metal and metal-boron interactions are less significant than the
$p-p$ covalent interaction of boron atoms. The bonding nature, elastic
property and hardness were investigated by Zhang {\it et al.}
\cite{ZLH+08} for HfB$_2$ as well as ZrB$_2$ using the first
principles total-energy plane-wave pseudopotential (PW-PP)
method. They also reported the elastic anisotropy, Poisson’s ratio,
hardness and Debye temperature in HfB$_2$ and ZrB$_2$. Deligoz {\it et
  al.} \cite{DCC10a} investigated the structural parameters (the
lattice constants and bond length) and phonon dispersion relations in
HfB$_2$ and TaB$_2$ compounds using the first-principles total energy
calculations. The secondary results on the temperature-dependent
behavior of thermodynamical properties such as entropy, heat capacity,
internal energy, and free energy were also presented. Zhang {\it et
  al.} \cite{ZLLH+10} investigated the ideal tensile and shear
strengths of TiB$_2$, ZrB$_2$ and HfB$_2$ by first-principles
stress-strain calculations. Due to the nonlinearity of the stress
response at large stains, the plastic anisotropy cannot be derived
from elastic constants. Based on the relative stiffness of boron
hexagons, a bond length indicator was obtained to characterize the
preference for basal or prismatic slip in diborides. Zhang {\it et
  al.} \cite{ZC10,ZCL11} investigated theoretically the pressure
dependence of elastic constants, bulk modulus and elastic anisotropy
of HfB$_2$. The pressure dependence of structural property shows that
the effect of pressure is little on the structure of HfB$_2$. They
find high pressure greatly changes the profile of the density of
states (DOS), but it hardly changes the DOS value at Fermi
level. Meanwhile, the Mulliken population analyses are
investigated. It was suggested that as the pressure increases, a
number of charge transfer from Hf to B atoms. Through quasi-harmonic
Debye model, the variations of the Debye temperature, heat capacity
and thermal expansion with pressure and temperature were obtained and
discussed.

Fedorchenko and Grechnev with coauthors \cite{FGP+09,GFL+09} measured
the temperature dependences of the magnetic susceptibility $\chi$ and
its anisotropy $\Delta \chi= \chi_{\parallel} - \chi_{\perp} $ for
single crystals of transition-metal diborides MB$_2$ (M = Sc, Ti, V ,
Zr, Hf) in the temperature interval 4.2 - 300 K. A transition into the
superconducting state was not found in any of the diborides studied,
right down to liquid-helium temperature. It was found that the
anisotropy is weakly temperature-dependent, a nonmonotonic function of
the filling of the hybridized $p-d$ conduction band. First-principles
calculations of the electronic structure of diborides and the values
of the paramagnetic contributions spin and Van Vleck to their
susceptibility show that the behavior of the magnetic anisotropy is
due to the competition between Van Vleck paramagnetism and orbital
diamagnetism of the conduction electrons. Authors of
Ref. \onlinecite{WOCB+04} determined the thermal conductivity, thermal
expansion, Young's modulus, flexural strength, and brittle-plastic
deformation transition temperature for HfB$_2$ as well as for
HfC$_{0.98}$, HfC$_{0.76}$, and HfN$_{0.92}$ ceramics. The thermal
conductivity of modified HfB$_2$ exceeded that of the other materials
by a factor of 5 at room temperature and by a factor of 2.5 at 820
$^{\circ}$C. The transition temperature of HfB$_2$ was 1100
$^{\circ}$C. Pure HfB$_2$ was found to have a strength of 340 MPa in 4
point bending, that was constant from room temperature to 1600
$^{\circ}$C, while a HfB2 + 10\% HfC$_x$ had a higher room temperature
bend strength of 440 MPa, but that dropped to 200 MPa at 1600
$^{\circ}$C. The results of the theoretical modeling suggest that
HfB$_2$ should survive the high thermal stresses generated during the
nozzle test primarily because of its superior thermal
conductivity. Yang {\it et al.}  \cite{YJSK+07} used {\it in situ}
spectroscopic ellipsometry to analyze HfB$_2$ thin films. By modeling
the film optical constants with a Drude-Lorentz model, the film
thickness, surface roughness, and electrical resistivity were
measured. By modeling the real-time data in terms of film thickness
and surface roughness, the film nucleation and growth morphology were
determined as a function of substrate type, substrate temperature, and
precursor pressure. Li {\it et al.} \cite{LZZC11} studied the
thermodynamics of the oxidation of HfB$_2$ at temperatures of 1000,
1500, 2000, and 2500 K using volatility diagrams. They found that
HfB$_2$ exhibits oxidation behavior similar to ZrB$_2$. Zhang {\it et
  al.} \cite{ZPMG11} investigated experimentally the thermal and
electrical transport properties of various spark plasma-sintered
HfB$_2$ and ZrB$_2$ based polycrystalline ceramics over the 298-700 K
temperature range. Measurements of thermal diffusivity, electrical
resistivity, and Hall coefficient were reported, as well as the
derived properties of thermal conductivity, charge carrier density,
and charge carrier mobility. Hall coefficients were negative
confirming electrons as the dominant charge carrier. A Wiedemann-Franz
analysis confirms the dominance of electronic contributions to heat
transport. The thermal conductivity was found to decrease with
increasing temperature. The properties of the Fermi surface of
ScB$_2$, ZrB$_2$, and HfB$_2$ single crystals were studied by
Pluzhnikov {\it et al.} \cite{PSD+07} using the de Haas-van Alphen
effect. The angular dependences of the frequencies of the dHvA
oscillations in the planes ($10\bar10$), ($11\bar20$), and (0001) and
the values of their effective cyclotron masses were measured.  The
frequencies of the oscillations found lie in the interval $(0.96-0.87)
\times 10^2$ T and the measured cyclotron masses $m_c^*$ lie in the
range $(0.09-0.87)m_0$.

Despite a lot of publications, there are still many open questions
related to the electronic structure and physical properties of HfB$_2$
diboride. The most theoretical efforts were devoted to the lattice and
mechanical properties of HfB$_2$. There is no theoretical expalanation
of the Fermi surface as well as angle dependence of the cyclotron
masses and extremal cross sections of the Fermi surface,
electron-phonon interaction and electrical resistivity in HfB$_2$. The
aim of this work is a complex investigation of the electronic
structure, Fermi sutface, angle dependence of the cyclotron masses and
extremal cross sections of the Fermi surface, phonon spectra,
electron-phonon Eliashberg and transport spectral functions, and
temperature dependence of electrical resistivity of the HfB$_2$
diboride. The paper is organized as follows. Section \ref{sec:details}
presents the details of the calculations. Section \ref{sec:results} is
devoted to the electronic structure as well as the Fermi surface,
angle dependence of the cyclotron masses and extremal cross sections
of the Fermi surface, phonon spectra, electron-phonon interaction and
electrical resistivity using the fully relativistic and full potenrial
LMTO band structure methods. The results are compared with available
experimental data. Finally, the results are summarized in
Sec.~\ref{sec:summ}.
  
\begin{figure}[tbp!] 
\begin{center} 
  \includegraphics[width=0.90\columnwidth]{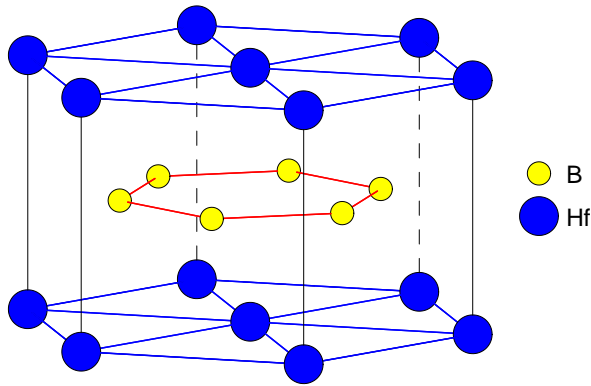} 
\end{center} 
\caption{\label{struc}(Color online) Schematic representation of the HfB$_2$
  crystal structure. }
\end{figure} 
 
\section{\label{sec:details}Computational details} 
 
Most known transition-metal (M) diborides MB$_2$ are formed by group III-VI
transition elements (Sc, Ti, Zr, Hf, V, Nb, and others) and have a layered
hexagonal C32 structure of the AlB$_2$-type with the space group symmetry
$P6/mmm$ (number 191). It is simply a hexagonal lattice in which
closely-packed transition metal layers are present alternative with
graphite-like B layers (Fig.~\ref{struc}). These diborides cannot be exactly
layered compounds because the inter-layer interaction is strong even though
the M layers alternate with the B layers in their crystal structure. The boron
atoms lie on the corners of hexagons with the three nearest neighbor boron
atoms in each plane. The M atoms lie directly in the centers of each boron
hexagon, but midway between adjacent boron layers. Each transition metal atom
has twelve nearest neighbor B atoms and eight nearest neighbor transition
metal atoms (six are on the metal plane and two out of the metal plane). There
is one formula unit per primitive cell and the crystal has simple hexagonal
symmetry ($D6h$). By choosing appropriate primitive lattice vectors, the atoms
are positioned at Hf (0,0,0), B ($\frac{1}{3}, \frac{1}{6}, \frac{1}{2}$), and
B ($\frac{2}{3}, \frac{1}{3}, \frac{1}{2}$) in the unit cell. The distance
between Hf-Hf is equal to $c$. This structure is quite close packed, and can
be coped with efficiently and accurately by the atomic sphere approximation
method. However, for precise calculation of the phonon spectra and
electron-phonon interaction, a full potential approximation should be used.
 
The Eliashberg function (the spectral function of the electron-phonon 
interaction) expressed in terms of the phonon linewidths 
$\gamma_{\mathbf{q}\nu}$ has the form \cite{Allen72} 
 
\begin{equation} 
\alpha^2F(\omega) = \frac{1}{2\pi N(\epsilon_F)}\sum_{\mathbf{q}\nu} 
\frac{\gamma_{\mathbf{q}\nu}}{\omega_{\mathbf{q}\nu}}\delta(\omega 
-\omega_{\mathbf{q}\nu}). 
\label{mu_Bgr} 
\end{equation} 
 
The line-widths characterize the partial contribution of each phonon: 
 
\begin{equation} 
\gamma_{\mathbf{q}\nu}= 2\pi\omega_{\mathbf{q}\nu}\sum_{jj' 
  \mathbf{k}} | g_{ \mathbf{k}+\mathbf{q}j', 
  \mathbf{k}j}^{\mathbf{q}\nu} |^2 \delta (\epsilon _{j \mathbf{k}} - 
\epsilon_{F}) \delta (\epsilon _{\mathbf{k}+\mathbf{q}j'} - 
\epsilon_{F}). 
\label{nu} 
\end{equation} 
 
The electron-phonon interaction constant is defined as: 
 
\begin{equation} 
\lambda_{e-ph} = 2\int_0^\infty\frac{d\omega}{\omega}{\alpha^2}F(\omega). 
\label{lamda} 
\end{equation} 
 
It can also be expressed in terms of the phonons line-widths: 
 
\begin{equation} 
\lambda_{e-ph} = \sum_{ \mathbf{q}\nu}\frac{\gamma_{ {\bf q}\nu}}{\pi N
  (\epsilon_F) \omega_{ {\bf q}\nu}^2},
\label{lamda2} 
\end{equation} 
were N($\epsilon_F$) is the electron density of states per atom and per spin
on the Fermi level ($\epsilon_F$) and $g_{ \mathbf{k}+ {\bf q}j'
\mathbf{k}j}^{ {\bf q}\nu}$ is the electron-phonon interaction matrix element.
The double summation over Fermi surface in Eq.(\ref{nu}) was carried out on
dense mesh (793 point in the irreducible part of the BZ)
 
Calculations of the electronic structure and physical properties of the
HfB$_2$ diborides were performed using fully relativistic LMTO method
\cite{APS+95} with the experimentally observed lattice constants: $a$=3.141
\AA\, and $c$=3.47 \AA\, for HfB$_2$. \cite{StPe75} For the calculation of the
phonon spectra and electron-phonon interaction a scalar relativistic FP-LMTO
method \cite{Sav96} was used. In our calculations we used the Perdew-Wang
\cite{PW92} parameterization of the exchange-correlation potential in general
gradient approximation. BZ integrations were performed using the improved
tetrahedron method. \cite{BJA94} Phonon spectra and electron-phonon matrix
elements were calculated for 50 points in the irreducible part of the BZ using
the linear response scheme developed by Savrasov. \cite{Sav96} The 5$s$ and
5$p$ semi-core states of HfB$_2$ were treated as valence states in separate
energy windows. Variations in charge density and potential were expanded in
spherical harmonics inside the MT sphere as well as 2894 plane waves in the
interstitial area with 88.57 Ry cut-off energy for HfB$_2$. As for the area
inside the MT spheres, we used 3k$-spd$ LMTO basis set energy (-0.1, -1, -2.5
Ry) with one-center expansions inside the MT-spheres performed up to $l_{max}$
= 6.

\section{\label{sec:results}Results and discussion} 
 
\subsection{Energy band structure} 
 
\begin{figure}[tbp!] 
\begin{center} 
\includegraphics[width=0.45\textwidth]{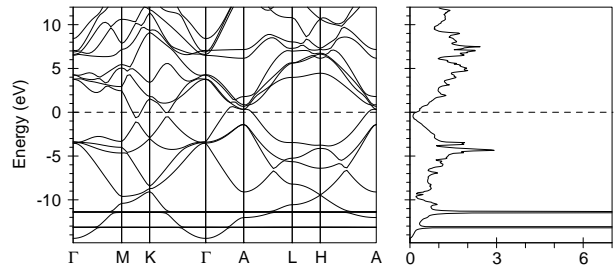} 
\end{center} 
\caption{\label{Ek} (Color online) Energy band structure and total DOS [in
  states/(cell eV)] of HfB$_2$.}
\end{figure} 

\begin{figure}[tbp!] 
\begin{center} 
\includegraphics[width=0.45\textwidth]{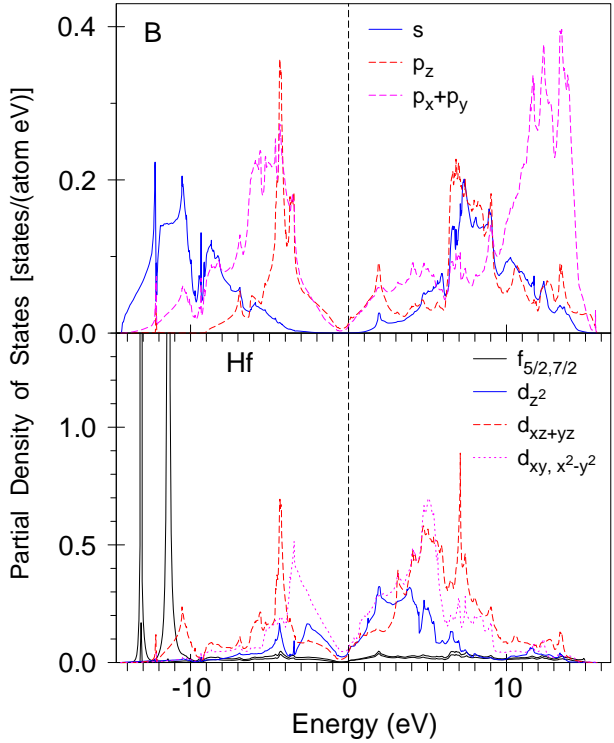} 
\end{center} 
\caption{\label{PDOS} (Color online) Partial DOSs [in states/(atom 
    eV)] of HfB$_2$.} 
\end{figure} 
 
Figure~\ref{Ek} presents the energy band structure and total density
of states of HfB$_2$. The partial DOSs HfB$_2$ are shown in
Fig. \ref{PDOS}. Our results for the electronic structure of HfB$_2$
are in agreement with earlier calculations.
\cite{FGP+09,DCC10a,FTH+10} A common feature for all transition metal
diborides is the deep DOS minimum (pseudo-gap) at the Fermi energy
separating the valence band and the conduction band. According to
Pasturel {\it et al.}, \cite{PCH85} a pseudo-gap arises because of a
strong chemical interaction. The M-B covalent bonding is believed to
be responsible for this effect. The Hf 4$f_{\frac{5}{2},\frac{7}{2}}$
states in HfB$_2$ are situated at the $-$14.5 eV to $-$10 eV. The Hf
5$d$ states are the dominant features in the interval from $-$12.5 eV
to 14 eV. These tightly bound states show overlap with B 2$p$ and, to
a lesser extent, with B 2$s$ states both above and below $\ef$,
implying considerable covalency. Higher-energy states between 9 eV and
17 eV above $\ef$ appear to arise from Hf 6$p$ and 6$s$ states
hybridized with B 2$p$ states. The crystal field at the Hf site ($D6h$
point symmetry) causes the splitting of Hf $d$ orbitals into a singlet
$a_{1g}$ ($\dzz$) and two doublets $e_{1g}$ ($\dyz$ and $\dxz$) and
$e_{2g}$ ($\dxy$ and $\dxxyy$). The crystal field at the B site ($D3h$
point symmetry) causes the splitting of B $p$ orbitals into a singlet
$a_4$ ($p_z$) and a doublet $e_2$ ($p_x$ and $p_y$). B $s$ states
occupy a bottom of valence band between $-$14.6 eV and $-$3.0 eV and
hybridize strongly with B $p_x$ and $p_y$ and Hf $\dyz$ and $\dxz$
states located at $-$12.5 eV to $-$0.5 eV. B $p_x$ and $p_y$ occupied
states are located between $-$12.5 eV and $-$0.5 eV.  B $p_z$ states
occupied a smaller energy interval from $-$7.5 eV to $-$0.5 eV with a
very strong and narrow peak structure at around $-$4 eV.

\subsection{Fermi surface} 
 
The Fermi surfaces (FS) of ScB$_2$, HfB$_2$ and HfB$_2$, were studied by
Pluzhnikov {\it et al.} \cite{PSD+07} using the dHvA effect.
 
\begin{figure}[tbp!] 
\begin{center} 
\includegraphics[width=0.45\textwidth]{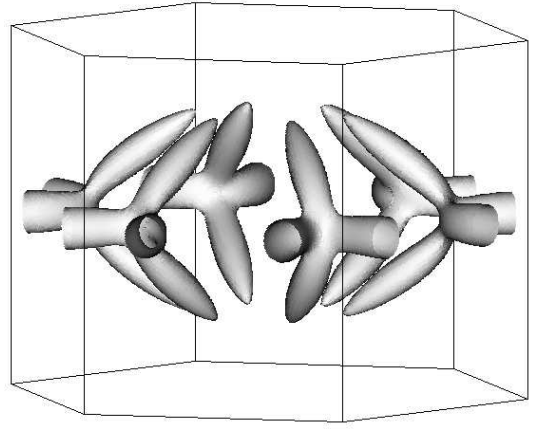} 
\end{center} 
\caption{\label{FS_el} (Color online) The calculated electron sheets of 
  the Fermi surface around K symmetry point from the 6th energy band 
  of HfB$_2$.} 
\end{figure} 
 
\begin{figure}[tbp!] 
\begin{center} 
\includegraphics[width=0.45\textwidth]{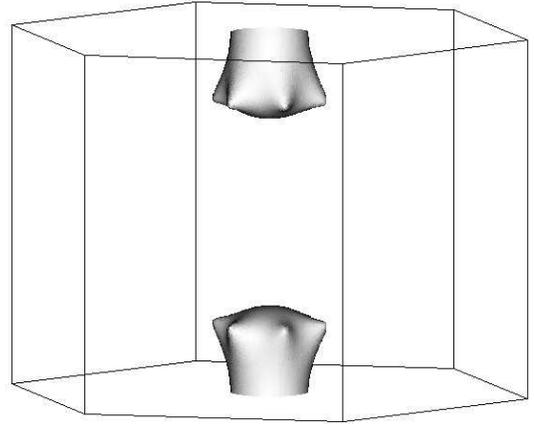} 
\end{center} 
\caption{\label{FS_hole} The calculated hole sheets of the Fermi surface 
  at the A symmetry point from the 5th energy band of HfB$_2$.} 
\end{figure}

\begin{figure}[tbp!] 
\includegraphics[width=0.4\textwidth]{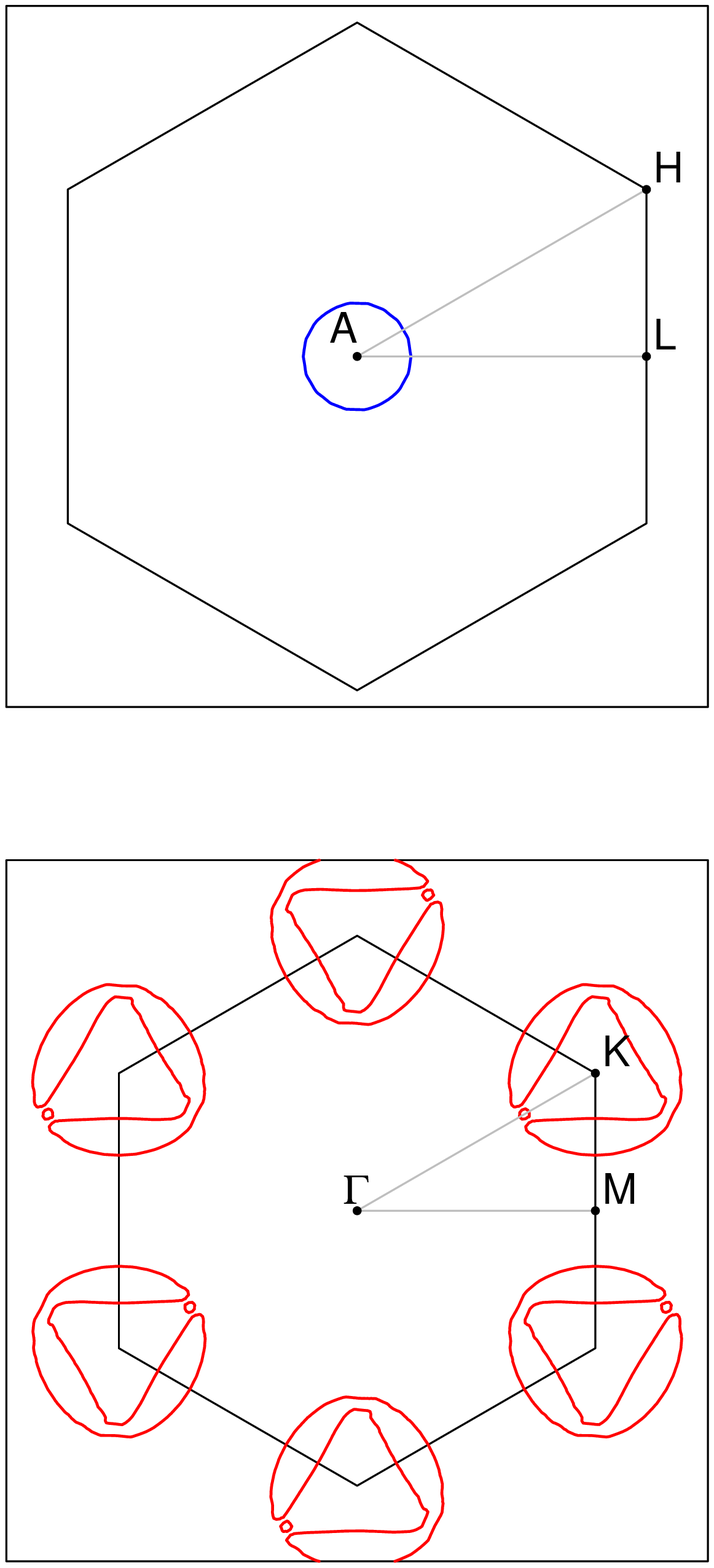} 
\caption{\label{FS_cs} (Color online) The calculated cross sections in the
  plane perpendicular $z$ direction and crossed $A$ symmetry point (upper
  panel) and $\Gamma$ point (lower panel) for HfB$_2$ (full red curves). }
\end{figure}

Theoretical calculations show a ring-like electron FS around the $K$
symmetry point (Fig. \ref{FS_el}) and of a wrinkled dumbell-like hole
FS at the A point (Fig. \ref{FS_hole}) in HfB$_2$. The electron and
hole Fermi surfaces have threefold and sixfold symmetries,
respectively. Figure \ref{FS_cs} shows the calculated cross section
areas in the plane perpendicular $z$ direction and crossed $A$
symmetry point for hole FS (upper panel) and crossed $\Gamma$ point
for electron FS (lower panel) of HfB$_2$.
 
\begin{figure}[tbp!] 
\begin{center} 
\includegraphics[width=0.45\textwidth]{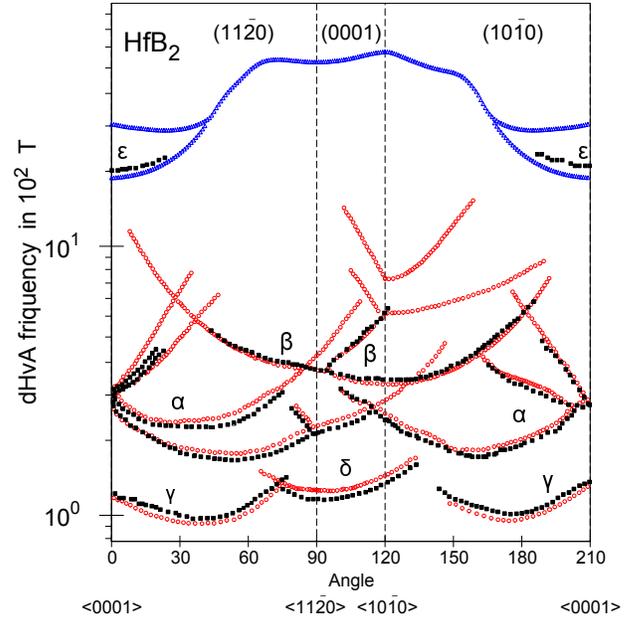} 
\end{center} 
\caption{\label{FS_Hf_S} (Color online) The calculated (open red and blue
  circles for the electron and hole surfaces, respectively) and experimentally
  measured \protect\cite{PSD+07} (black full squares) angular dependence of
  the dHvA oscillation frequencies in the compound HfB$_2$.}
\end{figure}

\begin{figure}[tbp!] 
\begin{center} 
\includegraphics[width=0.45\textwidth]{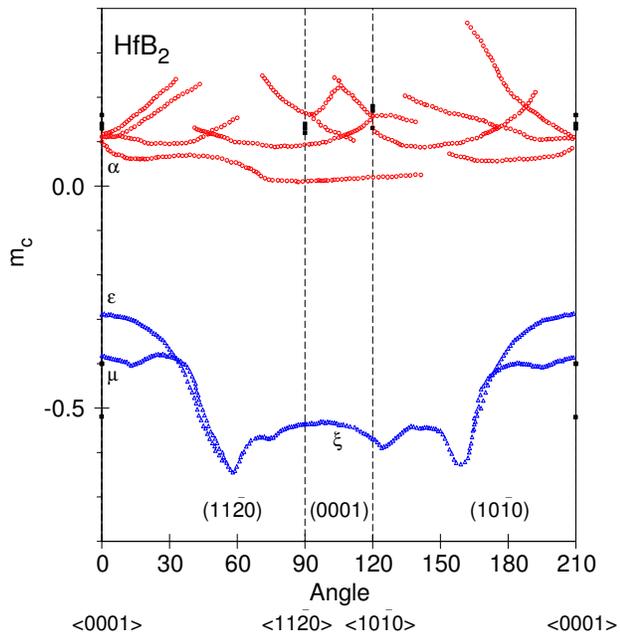} 
\end{center} 
\caption{\label{FS_Hf_mc} (Color online) The calculated angular dependence of
  the cyclotron masses for the electron Fermi surface (open red circles) and
  the hole Fermi surface (blue open triangles) and experimentally measured
  ones \protect\cite{PSD+07} (black full squares) in the compound HfB$_2$. }
\end{figure}

Figure \ref{FS_Hf_S} represents angular variations of the
experimentally measured dHvA frequencies \cite{PSD+07} for HfB$_2$ in
comparison with the first-principle calculations for field direction
in the ($10\bar10$), ($11\bar20$), and (0001) planes. The observed
frequencies of $\alpha$, $\beta$, $\gamma$, and $\delta$ oscillations
belong to electron FS around the $K$ point. The $\epsilon$, $\mu$, and
$\zeta$ orbits belong to the hole wrinkled dumbbell FS. The $\alpha$
frequencies have four branches at the ($10\bar10$) plane and three
branches at the ($11\bar20$) plane. The lower $\gamma$ frequencies
have one branch in both the planes. The theory reasonably well
reproduces the frequencies measured experimentally. However, there are
still some discrepancies. The $\beta$ orbits have an additional two
branches at higher frequencies at the ($11\bar20$), and (0001) planes
not observed experimentally. The experiment for high frequencies
detected only $\epsilon$ orbits in vicinity of the $<0001>$ direction
in HfB$_2$. We found the $\epsilon$, $\mu$ and $\zeta$ orbits similar
to the corresponding orbits observed experimentally in isostructural
and isovalent ZrB$_2$. \cite{PSD+07} These orbits have not been
detected in the dHvA experiment. \cite{PSD+07} One of the possible
reasons for that is the relatively large cyclotron masses for these
orbits. Figure \ref{FS_Hf_mc} shows the theoretically calculated
angular dependence of the cyclotron masses ($m_b$) and the
experimentally measured masses ($m_c^*$) for high symmetry directions
in HfB$_2$. The cyclotron effective masses were determined from the
temperature dependences of the amplitudes of the dHvA
oscillations. The cyclotron masses for the $\epsilon$, $\mu$, and
$\zeta$ orbits in HfB$_2$ are much higher than the corresponding
low-frequency oscillations $\alpha$, $\beta$, $\gamma$ and
$\delta$. The fact that the masses for electron Fermi surface are
significantly larger than for the hole Fermi surface may explain a
negative experimentally measured Hall coefficient \cite{ZPMG11} and
confirms electrons as the dominant charge carriers in HfB$_2$. A
Wiedemann-Franz analysis also indicate the dominance of electronic
contributions to heat transport. \cite{ZPMG11}

We note that band cyclotron effective masses $m_b$ are renormalized by
the electron-phonon interaction $m_c^*=m_b(1+ \lambda)$, where
$\lambda$ is the constant of the electron-phonon interaction. By
comparing the experimentally measured cyclotron masses with band
masses we can estimate the $\lambda$. It is strongly varied on the
orbit type and magnetic direction. We estimate the constant of the
electron-phonon interaction to be equal to 0.18-0.23 for the $\alpha$
orbits and 0.36 and 0.75 for the $\epsilon$ and $\mu$ orbits,
respectively, with H$\parallel <0001>$. For the $<10\bar10>$ and
$<11\bar20>$ directions the $\lambda$ for the $\alpha$ orbits are
reduced, respectively, to 0.10 and 0.12 values.

\subsection{Phonon spectra} 
 
\begin{figure}[tbp!] 
\begin{center} 
\includegraphics[width=0.45\textwidth]{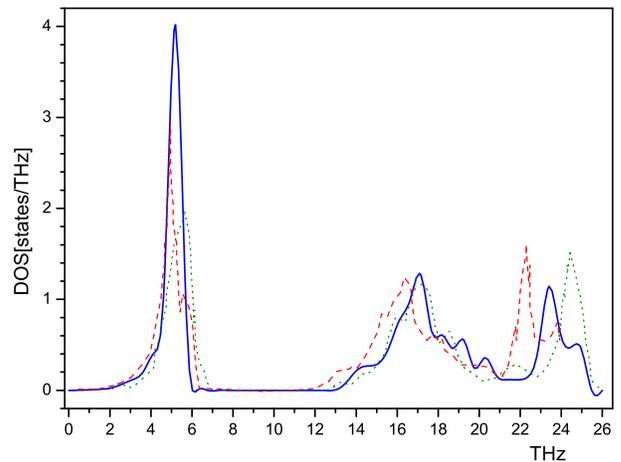} 
\end{center} 
\caption{\label{P_DOS} (Color online) Theoretically calculated phonon
  density of states (full blue line) for HfB$_2$. The dotted green and
  red dashed lines present the calculated phonon DOS of HfB$_2$ by
  Deligoz {\it et al.} \protect\cite{DCC10a} and Lawson {\it et al.}
  \protect\cite{LBD11}, respectively. }
\end{figure}

The unit cell of HfB$_2$ contains three atoms, which gives in general
case a nine phonon branches. Figure \ref{P_DOS} shows theoretically
calculated phonon density of state for HfB$_2$ (full blue curve). The
DOS for HfB$_2$ can be separated into three distinct regions. Based on
our analysis of relative directions of eigenvectors for each atom in
unit cell, we find that the first region (with a peak in phonon DOS at
5.2 THz) is dominated by the motion of Hf. This region belongs to the
acoustic phonon modes. The second wide region (14-20 THz) results from
the coupled motion of Hf and the two B atoms in the unit cell. The
$E_{1u}$, $A_{2g}$, $B_{1g}$ phonon modes (see Table \ref{fre}) lie in
this area. The phonon DOS in the third region extends from 22 THz to
26 THz. This is due to the movement of boron atoms and is expected
since boron is lighter than Hf. The covalent character of the B-B
bonding is also crucial for the high frequency of phonons. The
in-plane E$_{2g}$ mode belongs to this region. The second and third
regions represent optical phonon modes in crystals. The most
significant feature in the phonon DOS is a gap around 6-13 ThZ. This
gap is a consequence of the large mass difference between B(10.8 a.u.)
and Hf (178.49 a.u.), which leads to decoupling of the transition
metal and boron vibrations.
 
\begin{table}[tbp!] 
  \caption{\label{fre} Theoretically calculated phonon frequencies (in
    THz) in the $\Gamma$ symmetry point for HfB$_2$ and calculated
    phonon frequencies by Deligoz {\it et al.}
    \cite{DCC10a} and Lawson {\it et al.} Ref.
    \onlinecite{LBD11}. }
  
\begin{tabular}{|c|c|c|c|c|c|} 
\hline 
reference  & $E_{1u}$ & $A_{2g}$ & $B_{1g}$ & $E_{2g}$ \\ 
\hline 
 our results  & 13.76 & 15.03 & 17.12 & 25.17 \\ 
 SIESTA\cite{DCC10a} & 14.10 & 15.19 & 15.87 & 24.49  \\  
 VASP\cite{LBD11} & 13.34 & 14.00 & 16.40 & 24.16  \\  
 ABINIT\cite{LBD11} & 12.92 & 13.85 & 16.01 & 23.59  \\  
		         
\hline 
 
\end{tabular} 
\end{table} 
 
Currently, there are no data concerning the experimentally measured
phonon DOS in HfB$_2$. So we compare our results with theoretically
calculated phonon DOS by Deligoz {\it et al.} \cite{DCC10a} and Lawson
{\it et al.} \cite{LBD11} (Fig. \ref{P_DOS} and Table
\ref{fre}). Calculations of Deligoz {\it et al.} \cite{DCC10a} were
based on the density functional formalism and generalized gradient
approximation. They used the Perdew-Burke-Ernzerhof functional
\cite{PBE96} for the exchange-correlation energy as it is implemented
in the SIESTA code. \cite{OAS96} This code calculates the total
energies and atomic Hellmann-Feynman forces using a linear combination
of atomic orbitals as the basis set. The basis set consists of finite
range pseudoatomic orbitals of the Sankey-Niklewsky type \cite{SaNi89}
generalized to include multiplezeta decays. The interactions between
electrons and core ions are simulated with the separable
Troullier-Martins \cite{TrMa91} normconserving pseudopotentials. In
other words, they used the so-called "frozen phonon" technique and
built an optimized rhombohedral supercell with 36 atoms. This method
is inconvenient for calculating phonon spectra for small {\bf
  q}-points as well as for compounds with large number of atoms per
unit cell. Lawson {\it et al.} \cite{LBD11} used two different codes
to calculated the phonon spectra. VASP, the supercell method, based un
the projected augmented wave potentials.  Second method, ABINIT, used
Fritz Haber Institute pseudopotentials in the Troulliers-Martin
form. VASP results of Lawson {\it et al.}  \cite{LBD11} is slightly
closer to our calculation with respect to ABINIT data. There is a good
agreement between our calculations and the results of Deligoz {\it et
  al.}  \cite{DCC10a} in a shape and energy position of the second
peak in the phonon DOS. There is an energy shift towards smaller
energies of the first and third peaks of the Lawson {\it et al.}
\cite{LBD11} calculations in comparison with the Deligoz {\it et al.}
\cite{DCC10a} data with our results are just in between these two
calculations.

\subsection{Electron-phonon interaction} 
 
\begin{figure}[tbp!] 
\begin{center} 
\includegraphics[width=0.45\textwidth]{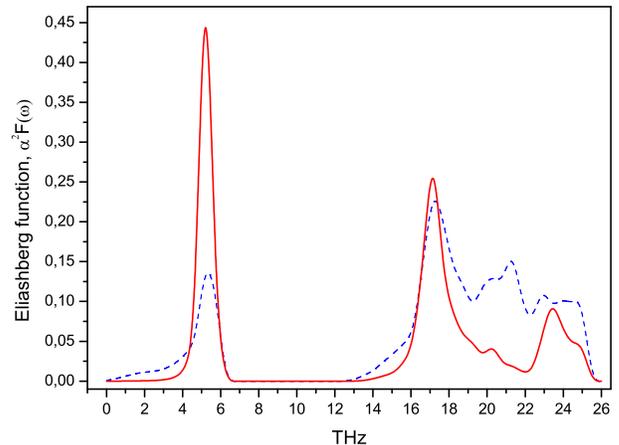} 
\end{center} 
\caption{\label{Eli} (Color online) Theoretically calculated
  Eliashberg function $\alpha^2 F(\omega)$ of HfB$_2$ (full red line)
  and electron-phonon prefactor $\alpha^2(\omega)$ (dashed blue
  line). }
\end{figure}

Figure \ref{Eli} shows theoretically calculated Eliashberg functions
for HfB$_2$ as well as electron-phonon prefactor $\alpha^2(\omega)$
(definition of this function is merely ratio
$\alpha^2(\omega)F(\omega)/F(\omega))$ . There is no difference
between main peaks positions of phonon spectra and electron-phonon
coupling function. Electron-phonon prefactor has three peaks: 5.2 THz,
17.1 THz and 21.3 THz (the corresponding peaks in the phonon DOS are
situated at the 5.2 THz, 17.1 THz, and 23.4 THz frequencies). The
$\alpha^2(\omega)$ has strongly varying character.  Therefore the
electron-phonon coupling can not be factorized into independent
electronic and phonon parts. The matrix element of electron-phonon
interaction cannot be represented in form $\alpha^2(\omega)\approx
const$ and hence well known McMillan approximation \cite{McMil68} is
not valid for HfB$_2$. By integrating the Eliashberg function we
estimate the averaged electron-phonon constants
$\lambda_{e-ph}$=0.17. The constant of the electron-phonon interaction
also can be estimated by comparison the theoretically calculated DOS
at the Fermi level with the electron specific heat coefficient
$\gamma$. $C_p=\gamma T$, where $\gamma$ =1.0 mJ·mole$^{-1}$K$^{-2}$
for HfB$_2$. \cite{TTC69} HfB$_2$ possesses quite small value of the
DOS at the Fermi level of 0.4 states/(cell eV), it gives the
theoretically calculated $\gamma_b$=0.8 mJ·mole$^{-1}$K$^{-2}$ and
$\lambda$=0.2 with qualitative agreement with $\lambda_{e-ph}$=0.17.

\subsection{Electrical resistivity} 
 
In the pure metals (excluding low-temperature region), the electron-phonon 
interaction is the dominant factor governing electrical conductivity of the 
substance. Using lowest-order variational approximation, the solution for the 
Boltzmann equation gives the following formula for the temperature dependence 
of $\rho_{I}(T)$: 
 
\begin{equation} 
\rho_I (T) = \frac{\pi\Omega_{cell} k_B T}{N(\epsilon_F)\langle
  v_I^2\rangle}\int_0^\infty\frac{d\omega}{\omega}\frac{\xi^2}{sinh^2\xi}{\alpha_{tr}^2}F(\omega),
\label{resist} 
\end{equation} 
where, the subscript $I$ specifies the direction of the electrical current. In
our work, we investigate two direction: [0001] (c-axis or z direction) and
[10$\bar1$0] (a-axis or x-direction).  $\langle v_I^2\rangle$ is the average
square of the $I$ component of the Fermi velocity, $\xi=\omega/{2k_BT}$.
 
Mathematically, the transport function $\alpha_{tr} F(\omega)$ differs from 
$\alpha F(\omega)$ only by an additional factor 
$[1-v_{I}(\mathbf{k})v_{I}(\mathbf{k}')/\langle v_I^2\rangle]$, which 
preferentially weights the backscattering processes. 
 
Formula (\ref{resist}) remains valid in the range $\Theta_{tr}/5 < T <
2\Theta_{tr}$ \cite{Sav96} where:

\begin{equation} 
\Theta_{tr} \equiv \langle\omega^2\rangle_{tr}^{1/2} , 
\label{debay} 
\end{equation}

\begin{equation} 
\langle\omega^2\rangle_{tr}= 
\frac{2}{\lambda_{tr}}\int_0^\infty\omega\alpha_{tr}^2 
F(\omega)d\omega , 
\label{deb2} 
\end{equation}

\begin{equation} 
\lambda_{tr}=2 \int_0^\infty\alpha_{tr}^2 F(\omega)\frac{d\omega}{\omega} , 
\label{deb3} 
\end{equation} 
 
The low-temperature electrical resistivity is the result of
electron-electron interaction, size effects, scattering on impurities,
etc., however, for high temperatures it is necessarily to take into
account the effects of anharmonicity and the temperature smearing of
the Fermi surface. In our calculations $\Theta_{tr}$=654.4 K for
$c$-axis, and 679.9 for $a$-axis for HfB$_2$.
 
\begin{figure}[tbp!] 
\begin{center} 
\includegraphics[width=0.45\textwidth]{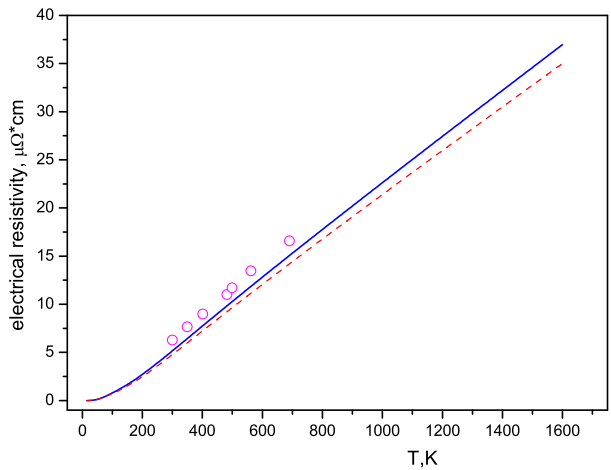} 
\end{center} 
\caption{\label{P3} (Color online) Theoretically calculated for the
  $<0001>$ direction (full blue curve) and the basal $<10\bar10>$
  direction (dashed red curve) and experimentally measured temperature
  dependence of electrical resistivity of
  HfB$_2$. \protect\cite{Gasch11} }
\end{figure} 
 
Figure \ref{P3} represents the theoretically calculated temperature
dependence of electrical resistivity of HfB$_2$ for the $<0001>$
direction (full blue curve) and the basal $<10\bar10>$ direction
(dashed red curve) and experimental measurements for polycrystalline
HfB$_2$.  \cite{Gasch11} Specimen of ceramic HB$_2$ was obtained by
spark plasma sintering method and had good ratio of experimental and
theoretically calculated density $\rho_{exp}/\rho_{th}=98.1\%$ Our
results are in good agreement with the experiment. The small
discrepancy does not exceed accuracy of calculation.  We obtained
anisotropy ratio of electrical resistivity at T=300K: $\rho_z/\rho_x$
= 1.079. Actually this fact indicates that for HfB$_2$ anisotropy is
not clearly expressed.

\section{\label{sec:summ}Summary} 
 
We have studied the electronic structure and physical properties of
HfB$_2$ using fully relativistic and full potential linear muffin-tin
orbital methods. We study the electron and phonon subsystems as well
as the electron-phonon interaction in this compound.
 
We investigated the Fermi surface, angle dependence of the cyclotron
masses, and extremal cross sections of the Fermi surface of HfB$_2$ in
details. Theoretical calculations show a ring-like electron FS in
HfB$_2$ around the $K$ symmetry point and a wrinkled dumbbell-like
hole FS at the A point. Theory reproduces the experimentally measured
dHvA frequencies in HfB$_2$ reasonably well. We found that masses for
low-frequency oscillations $\alpha$, $\beta$, $\gamma$, and $\delta$
are less than 0.25$m_0$. Masses for high-frequency oscillations
$\epsilon$, $\mu$, and $\zeta$ lie in the range from $-$0.3 to $-$0.65
$m_0$. The experiment for high frequencies detected only $\epsilon$
orbits in vicinity of the $<0001>$ direction in HfB$_2$. We found the
$\epsilon$, $\mu$ and $\zeta$ orbits similar to the corresponding
orbits observed experimentally in isostructural and isovalent
ZrB$_2$. These orbits have not been detected in the dHvA
experiment. One of the possible reasons for that is the relatively
large cyclotron masses for these orbits.
 
Calculated phonon spectra and phonon DOSs for HfB$_2$ is in good agreement
with previous calculations. We did not found regions with high electron-phonon
interaction or phonon dispersion curves with soft modes in HfB$_2$. This is in
agreement with the fact that no trace of superconductivity was found in these
borides. The averaged electron-phonon interaction constant was found to be
rather small $\lambda_{e-ph}$=0.17 for HfB$_2$. We calculated the temperature
dependence of the electrical resistivity in HfB$_2$ in the lowest-order
variational approximation of the Boltzmann equation. We found rather small
anisotropic behavior of the electrical resistivity in HfB$_2$ to be in good
agreement with experimental observation.
 
\section*{Acknowledgments} 
 
This work was supported by the National Academy of Sciences of Ukraine
in the framework of the State Target Scientific and Technology Program
"Nanotechnology and Nanomaterials" for 2010-2014 (No. 0277092303) and
Implementation and Application of Grid Technologies for 2009-2013
(No. 0274092303).


\newcommand{\noopsort}[1]{} \newcommand{\printfirst}[2]{#1}
  \newcommand{\singleletter}[1]{#1} \newcommand{\switchargs}[2]{#2#1}

\end{document}